# Ownership Structure Variation and Firm Efficiency

Sallahuddin Hassan[1], Zalila Othman[1] & Mukaramah Harun[1]

[1] School of Economics, Finance and Banking, UUM COB, Universiti Utara Malaysia, Malaysia

Correspondence: Sallahuddin Hassan, School of Economics, Finance and Banking, UUM COB, Universiti Utara Malaysia, 06000 Sintok, Kedah, Malaysia. Tel: 60-13-528-1898. E-mail: din636@uum.edu.my

**Abstract**

Firms with different ownership structures could be argued to have different levels of efficiency. Highly concentrated firms are expected to be more efficient as this type of ownership structure may alleviate the conflict of interest between managers and shareholders. In Malaysia, public-listed firms have been found to have highly concentrated ownership structure. However, whether this evidence holds for every industry has not been established. Hence, the objective of this paper is to investigate whether there are variations in ownership structure and firm's efficiency across sectors. To achieve this objective, the frequency distributions of ownership structure were calculated and firms' efficiency scores for consumer products, industrial products, construction and trading/services sectors were measured. Data Envelopment Analysis (DEA) under the assumptions of constant returns to scale (*CRS*) and variable returns to scale (*VRS*) was employed to estimate firms' efficiency scores. A sample of 156 firms listed on the Kuala Lumpur Stock Exchange (KLSE) was selected using the stratified random sampling method. The findings have shown that there are variations in firm ownership structure and efficiency across sectors.

## 1. Introduction

Firm ownership structure in Malaysia has changed as business organizations flourish and the national economy grows. In addition, it has also changed due to economic transition, industrialization and privatization policies implemented by the government over the past four decades. As a result, size of firms have expanded and caused susbstantial changes in the dispersion, redistribution and concentration of ownership structure. Ramli (2010), for instance, has found that ownership structure in Malaysia is concentrated and large shareholders are in control. He found that the largest shareholders or a shareholder group owns around 40% of companies' paid-up capital. In addition, according to the study by Claessens, Djankov, Fan and Lang (2000) on corporations in nine East Asian countries, Malaysia has the third highest concentration of control after Thailand and Indonesia.

In general, firms with different ownership structure are argued to have different levels of efficiency. Highly concentrated firms will be more efficient because this type of ownership structure may alleviate the conflict of interests between managers and shareholders. This statement is mainly grounded in the very well-known principal-agency problem model by Jensen and Meckling (1976). The principal-agency problem argues that managerial share-ownership may reduce managerial incentives for privileges, expropriating shareholders' wealth or engaging in other sub-optimal activities. This subsequently helps in aligning the interest of managers and shareholders and consequently lowers agency costs and increases firm value. Regarding diffused shareholding, this kind of ownership structure does not provide adequate control to the shareholders due to the lack of capacity and motivation to monitor management decisions. Hence, the model predicts that larger managerial ownership stakes should lead to better firm performance.

The study by Abdul Samad (2002) reveals the existence of a high concentration of ownership in Malaysia. His study on the public-listed companies in Malaysia found that the means of shareholdings of the largest shareholders and the five largest shareholders to be about 30% and 60%, respectively. This finding indicates that companies on the Kuala Lumpur Stock Exchange (KLSE) are less diffused but are dominated by companies with concentrated shareholders, typically families or government-owned or promoted institutions.

Therefore, in view of the previous evidence of high ownership concentration of firms in Malaysia, this study attempts to investigate the variation in ownership structure and the variation in efficiency of the public-listed firms in Malaysia. By focusing our investigation on the variations in ownership structure and efficiency across sectors, we expect to gain a greater understanding of the actual scenario of firms' ownership structure and efficiency in Malaysia and to ascertain whether the efficiency of public listed firms in Malaysia differs across sectors. By conducting our analysis across sectors, this study hopes to add new and meaningful contributions to the previous literature on firms' efficiency and ownership structure.

The rest of this paper is organized as follows: the next section provides a brief literature review, followed by the methodology, discussion of results, policy implications and conclusion.

## 2. Literature Review

This review provides a brief discussion of the literature on ownership structure and firm efficiency and firm performance. A firm's ownership structure is an important driver of its efficiency and profitability (Church & Ware, 2000). A firm which operates at a high efficiency level can normally be associated with higher performance or profitability. Morck, Schleifer and Vishny (1988) have suggested that firm performance is likely to vary across industries.

In general, a firm's ownership structure can differ along two main dimensions. First, the degree of ownership concentration, which refers to whether ownership of a firm is more or less-dispersed. Second, the nature of the owners, which refers to who owns the majority stake in the firm. Firms' ownership structure can significantly influence firms' efficiency, such as through its influence on the decision making process. Specifically, a firm's ownership structure determines its capital structure. Furthermore, a firm's capital structure influences its investment and growth behavior. Both factors then become the major driving force of future returns to a firm and eventually affect firm efficiency.

Furthermore, various researchers such as Williamson (1964), Marris (1964), Galbraith (1967), Pfeffer and Salancik (1978), Salami (2011), Su and He (2012), Bejan and Bidian (2012) have argued that ownership structure has important implications for firm efficiency and strategic development. Based on the principal-agency problem model, Jensen and Meckling (1976) have argued that a firm's performance will be negatively related to low ownership concentration and positively related to high ownership concentration. This can be explained by the concept of diffused shareholding. Salami (2011) has discovered two important findings. First, most of the listed companies have high ownership concentration in structure; and second, corporate governance structure is an important element in the investment strategies of these companies. These two findings appear to show some positive relationship with corporate investment efficiency. Thus, the study shows that there exists a positive relationship amongst the share-ownership structure, corporate governance structure, and corporate profitability.

In contrast, Morck *et al*. (1988) believed that concentrated ownership may be associated negatively with firm performance where the overall effect on firm value may be positive at low concentration but negative at high concentration levels. Similarly, Su and He (2012) found similar results. Their study of 744 public-listed manufacturing firms in China for the period 1999 to 2006 found that firm efficiency, estimated using stochastic frontier analysis and data envelopment analysis, is negatively related to state ownership but positively related to public and employee share-ownership. They also found that the relationship between ownership concentration and firm efficiency is U-shaped, illustrating that the largest shareholder may engage in tunneling activities. As the identity of the largest shareholder changes from government or government-controlled legal entity to other types of legal entity, firm efficiency significantly improves. This indicates that firms with more independent board are more efficient. This evidence supports the idea that board of directors can be an effective internal governance mechanism.

In addition, the study by Iannotta, Nocera and Sironi (2006) have divided the European banking industry ownership classification into privately-owned banks (POBs), mutual banks (MBs), and government-owned banks (GOBs). Their findings have shown that ownership concentration does not significantly affect profitability. The study by Altanbas, Evans and Molyneux (2001) uses a variety of approaches to model cost and profit efficiencies for different ownership types in the German Banking market did not find strong evidence in support of the proposition that private ownership is associated with better firm performance and enterprise efficiency than state ownership. They found that privately-owned banks are less efficient than their mutual and public sector counterparts. Their inefficiency measures indicate that private banks have slight cost and profit disadvantages over their mutual and public sector competitors.

Meanwhile, another group of researchers such as Jensen and Meckling (1976), Demsetz (1983), Fama and Jensen (1983), and Demsetz and Lehn (1985) have claimed the irrelevancy of ownership structure. For instance, Demsetz (1983) argued that there should be no systematic relationship between variations in ownership structure and variations in firm performance. Thomsen and Pedersen (2000) have argued that the effect of concentrated or dispersed ownership on firm performance will be difficult to predict unless one controls for the firm's capital structure choice. In other words, the relationship between ownership concentration and firm performance also depends on the identity of the major shareholders. Different types of shareholders have different investment priorities and preferences for how to deal with principal-agency problem. Even though these differences can be solved in board meetings when consensuses are achieved, the implementation will be executed by the managers who might have different objectives from the shareholders. While shareholders' objective is to get high profit, managers' objectives may be getting high profit, good working environment, good salary, and improvement of workers' welfare.

Furthermore, other studies such as Ang, Hauser and Lauterbach (1997) and Lauterbach and Vaninsky (1999) for instance, have used the second criterion in classifying ownership structure. They distinguished between non-owner managed firms, firms controlled by concerns, firms controlled by a family, and firms controlled by a group of individuals (partners). The latter found that dispersed ownership and non-owner manager promotes firm performance, estimated as the actual net income of the firm divided by the optimal net income given by the firm's inputs. These findings can be supported by the studies of Morck *et al.* (1988) and McConnell and Servaes (1990), which have demonstrated that percentage ownership appears insufficient for describing the control structure. For instance, two firms with identical overall percentage ownership by large blockholders are likely to have different control organizations, depending on the identity of the large blockholders. They also argued that the relation between percentage ownership and firm performance is non-linear. Therefore, as stressed by Shleifer and Vishny (1997), the implication is that, it is important, not only how much equity a shareholder owns, but also who this shareholder is, that is, a private person, financial institution, non-financial institution enterprise, multi-national corporation or government.

In Malaysia, according to Thillainathan (1999), the controlling shareholder (i.e. those holding more than 50% ownership) through the pyramid structure is common. The controlling shareholders, either individuals/families or firms, are in the position to expropriate minority interests using their dominant voting right. Thillainathan's findings proved that large ownership or ownership concentration may contribute to deficiencies in corporate governance. Furthermore, the study by Gedajlovic and Shapiro (2002) on Japanese firms used both measures of ownership structure. Their first measure is ownership by five largest blockholders, which does not distinguish shareholders' identities. The second and third measures are ownership by financial institutions and ownership by non-financial institutions, where each measure is equal to the percentage of a company's outstanding shares held by Japanese financial and non-financial companies. Their results show a positive relationship between ownership concentration and financial performance, which is consistent with agency theory prediction. This finding suggests that large Japanese investors can operate as effective monitors of top executives in other firms. In addition, their results also indicate that shareholders identity matters. Their conclusion is based on the fact that redistribution effects (transfer of profits from profitable firm to less profitable firm) was found only when they examined the ownership of financial and non-financial firms. Such a result is consistent with the notion that distinct classes of shareholders differ in their investment objectives and capacities to influence corporate behavior (Thomsen & Pedersen, 2000).

Gedajlovic and Shapiro (2002) have followed Demsetz and Villalonga (2001) method. They modelled ownership structure, first, as an endogenous variable and second, they examined two different measures of ownership structure: (a) the fraction of shares owned by insiders (top management, CEO, board members) and (b) fraction of shares owned by important outside investors. Their findings indicate that there exists a linear positive relationship between profitability and ownership structure. Furthermore, their results suggests that the greater the degree to which shares are concentrated in the hands of outside or inside shareholders, the more effectively management behavior is monitored and disciplined, leading to better performance. In addition, their results prove the endogeneity of ownership structure, where profitability is a positive determinant of ownership structure. There exists evidence that superior firm performance leads to an increase in the value of stock options owned by management or large shareholders, which if exercised, would increase their share ownership.

Therefore, based on these different views and findings, there seems to be inconclusive evidence on the relationship between ownership structure and firm efficiency. This situation has captured our interest to investigate the issue further. In this study, specifically, we consider the effect of ownership concentration on firm performance, specifically on firm efficiency. In addition, since the majority of the studies on the relationship between ownership structure and firm performance have been on firms in the developed countries, our study aims to extend previous knowledge by examining this relationship in a developing country. Malaysia is a developing country which provides a rich setting for exploring this issue further. To the best of our knowledge, this kind of study has not been done extensively in Malaysia. Amongst the few studies on the issue in the Malaysian setting include Abdul Samad (2002), Haniffa and Hudaib (2006), Chang and Shazali (2005), and Faizah (2006).

## 3. Methodology

This study was carried out using secondary data from a sample of 156 firms which were listed on the KLSE. Unbalanced panel data of listed firms over the period 2000 through 2010 were used in our analysis. The sample firms were selected using the stratified random sampling method. Data were gathered from Datastream and also from published reports of the selected companies. The final sample consists of firms from the consumer products, industrial products, construction, trading/services and properties sectors. Other sectors were dropped because the required data on output and inputs were not available for all the years of the study.

Ownership structure is measured using three concentrations of equity ownership namely, one-ownership concentration ratio (*CR1*), two-ownership concentration ratio (*CR2*), and four-ownership concentration ratio (*CR4*). The calculations of these concentrations are based on the analysis of shareholdings section disclosed in the firms' annual reports. This operationalization of ownership concentration concept has been widely used in previous researches such as by McConnel and Servaes (1990), Leech and Leahy (1991), and Claessens *et al*. (1999). Meanwhile, firm efficiency score is measured by the non-parametric method, Data Envelopment Analysis (DEA). The DEA approach was chosen over the Stochastic Frontier Analysis (SFA) which is a parametric approach since the former does not require the functional form and the distribution type to be assumed in advance. The DEA was originally developed by Farrel (1957) and subsequently extended and reformulated by Charnes, Fooper and Rhodes (1978). The DEA method by Charnes *et al.* is considered suitable for this study since it measures the efficiency of a micro unit relative to the efficiency of all the other micro units, assuming that all micro units are on or below the frontier. In the DEA method, constant returns to scale (*CRS*) and variable returns to scale (*VRS*) were employed to estimate firms' efficiency scores. The estimated model for technical efficiency is illustrated by Equation (1).

$$\theta^* = min\,\theta \qquad (1)$$

*subject to :*

$$\sum_{j=1}^{n} \lambda_j x_{ij} \leq \theta x_{io} \quad i = 1,2,...,m$$

$$\sum_{j=1}^{n} \lambda_j y_{rj} \leq y_{ro} \quad r = 1,2,...,s$$

$$\sum_{j=1}^{n} \lambda_j = 1$$

$$\lambda_j \geq 0 \qquad j = 1,2,...,n$$

where $DMU_0$ represents one of the *n* DMUs under evaluation, and $x_{io}$ and $y_{ro}$ are the *i*th input and *r*th output for $DMU_0$, respectively. $\lambda_j$ are unknown weights, where $j = 1, 2, …n$ represents the number of DMUs. The optimal value or $\theta^*$ represents the distance of the firm from the efficient frontier. Therefore, the most efficient firm will have $\theta^* = 1$ and the inefficient firm will have $\theta^* < 1$.

One of the most important steps in the estimation of efficiency is in the selection of input and output variables to be used in the DEA model. Physical measures and monetary measures are common types of input/output variables. We used monetary measures for three reasons. First, it is difficult to obtain variable information in physical units. Second, following Battese and Coelli (1995), it is preferable to use monetary measures to measure efficiency at the firm level since a firm is often engaged in many different activities. Third, using monetary measures may capture more information compared to physical measures. Thus, consistent with Feroz, Kim and Raab (2003), we chose two conventional input variables (wages and salaries as a measure of labour expenses, expenses on land, building and equipment, and interest expenses and one conventional output variable (revenue) in our DEA model.

## 4. Discussion of Results

Four sectors, namely consumer products, industrial products, construction, and trading/services were selected as the unit of analysis. The number of firms in the sample is determined by the proportion of firms in each sector to the entire firms registered on the KLSE. Firms in the trading/services sector (58 firms or 37.2%) represent the highest proportion of firms in the sample, followed by the industrial products sector (57 firms or 36.5%), consumer products (29 firms or 18.6%), and construction (12 firms or 7.7%).

Our investigation of ownership structure data reveals that the concentration of equity ownership in the 156 public-listed firms varies widely. The outcomes of the investigation are shown in Table 1, Table 2 and Table 3. Table 1 shows that industrial products, construction and trading/services have the highest share controlled by the single largest shareholder at the percentage shareholder range of 11 – 30%. There are 69 firms (44.23%) of the total firms that fall within this range across the sectors. Meanwhile, the consumer products sector has the single largest shareholder in the percentage shareholder range of 31 – 50%, the second largest ownership range. The total number of firms in this percentage shareholder range is 15 firms (51.70%) of total firms in this sector.

Meanwhile, Table 2 shows that highest share percentage controlled by the top two largest shareholders falls within the 31 – 50% range. In general, industrial products and trading/services sectors have the highest percentage shareholder distribution in this range. The total number of firms in this range is 57 firms (36.54%). Of the total firms that fall within this percentage range, the industrial product sector represents 40.35% of the firms, construction 10.53%, and trading/ services 31.58%. The highest share percentage for consumer products sector is in the 51 – 70% range. There are 13 firms which fall in this category, which constitute 44.83% of firms in this sector.

Extending the shareholding analysis to the top four largest shareholders, the study found that 69 of the 156 firms or 44.23% of the sample firms have between 31 – 50% of shares controlled by the four largest shareholders (shown in Table 3). Only one firm, in trading/services has more than 90% of shares in the hands of the four largest shareholders.

Table 1. Frequency distribution of ownership structure (Share controlled by the single largest shareholders)

| Sector of Firms | Ownership (%) | | | | | Total |
|---|---|---|---|---|---|---|
| | ≤ 10 | 11 – 30 | 31 – 50 | 51 – 70 | 71 – 90 | |
| Consumer Products | 1 | 9 | 15 | 4 | 0 | 29 |
| Industrial Products | 2 | 28 | 20 | 7 | 0 | 57 |
| Construction | 1 | 8 | 3 | 0 | 0 | 12 |
| Trading/Services | 6 | 24 | 18 | 9 | 1 | 58 |
| Total | 10 | 69 | 56 | 20 | 1 | 156 |

Table 2. Frequency distribution of ownership structure (Share controlled by top two largest shareholders)

| Sector of Firms | Ownership (%) | | | | | Total |
|---|---|---|---|---|---|---|
| | ≤ 10 | 11 – 30 | 31 – 50 | 51 – 70 | 71 – 90 | |
| Consumer Products | 0 | 3 | 10 | 13 | 3 | 29 |
| Industrial Products | 1 | 16 | 23 | 14 | 3 | 57 |
| Construction | 0 | 4 | 6 | 2 | 0 | 12 |
| Trading/Services | 0 | 17 | 18 | 18 | 5 | 58 |
| Total | 1 | 40 | 57 | 47 | 11 | 156 |

Table 3. Frequency distribution of ownership structure (Share controlled by top four largest shareholders)

| Sector of Firms | Ownership (%) | | | | | Total |
|---|---|---|---|---|---|---|
| | 11 – 30 | 31 – 50 | 51 – 70 | 71 – 90 | > 90 | |
| Consumer Products | 1 | 9 | 15 | 4 | 0 | 29 |
| Industrial Products | 2 | 28 | 20 | 7 | 0 | 57 |
| Construction | 1 | 8 | 3 | 0 | 0 | 12 |
| Trading/Services | 6 | 24 | 18 | 9 | 1 | 58 |
| Total | 10 | 69 | 56 | 20 | 1 | 156 |

Furthermore, Table 4 shows the descriptive statistic of estimated efficiency score for all sectors. Comparisons across sectors are made so as to highlight the variation in scores across different industries within the KLSE. *EFF1INCRS* and *EFF1INVRS* refer to efficiency score assuming *CRS* and *VRS*, respectively. It is clear that the mean of estimated efficiency score of *EFF1INCRS* for trading/services, at 0.032, is the highest compared to the other sectors. The maximum value of estimated *EFF1INCRS* score for consumer products and trading/services recorded the extreme value of efficiency, or 1.000. At the same time, trading/services sector also recorded the lowest minimum value of estimated *EFF1INCRS* score. Meanwhile, the mean of estimated *EFF1INVRS* scores are between 0.187 and 0.222. Consumer products, industrial products and trading/services sectors recorded the highest maximum value of *EFF1INVRS* scores, or 1.000. The lowest minimum value of *EFF1INVRS* score is recorded by the consumer products sector. The comparison between sectors is made to gain a greater understanding of the actual scenario of firms' ownership structure and efficiency in Malaysia and to explore their variations across sectors. The difference in efficiencies may be due to the different ownership structure in each sector.

Table 4. Descriptive statistic of estimated efficiency score

| Variable | Sector of Firm | Mean | Std. Dev | Min | Max |
|---|---|---|---|---|---|
| EFF1INCRS | Consumer Products | 0.014 | 0.088 | 0.00057 | 1.000 |
| | Industrial Products | 0.017 | 0.062 | 0.00075 | 0.527 |
| | Construction | 0.016 | 0.018 | 0.00187 | 0.119 |
| | Trading/Services | 0.032 | 0.107 | 0.00005 | 1.000 |
| EFF1INVRS | Consumer Products | 0.190 | 0.179 | 0.004 | 1.000 |
| | Industrial Products | 0.222 | 0.191 | 0.008 | 1.000 |
| | Construction | 0.207 | 0.163 | 0.013 | 0.642 |
| | Trading/Services | 0.187 | 0.218 | 0.005 | 1.000 |

In summary, our findings have portrayed the variation in firm ownership structures and efficiencies across sectors. Based on Table 1, Table 2, and Table 3, either industrial products or trading/services has the top largest shareholders in each bracket of ownership concentration. These results are consistent with the estimates of efficiency score in Table 4 where industrial products and trading/services sectors have the highest score under CRS and VRS assumptions, respectively. Therefore, based on this evidence, it seems that ownership structure may influence firms' efficiency. In other words, a sector which has a high concentration of ownership may tend to be efficient.

In the present study, the distribution of ownership is classified into three categories: the share concentration of the single largest, the top two largest, and the top four largest shareholders in each sector. In the first category, the largest proportion of firms in the industrial products, construction, and trading/services have a share concentration in the 11 – 30% range. However, the largest number of firms in consumer products has a share concentration in the 31 – 50% range. The picture changes for the second and third category of ownership. The highest proportion of firms in the industrial products, construction, and trading/services have a share concentration in the 31 – 50% range. On the other hand, the largest proportion of consumer product firms has a share concentration in the 51 – 70% range.

## 5. Policy Implications

Based on the results of the analyses presented and discussed in the previous sections, this study has been able to provide a clearer picture of the relationship between firms' ownership structure and efficiency in Malaysia. The findings of this study provide support for the hypothesis that firm efficiency is largely affected by its pattern of ownership structure. Firm efficiency is caused mainly by high concentration of ownership structure. Since large shareholders are a common phenomenon in Malaysian firms, this would be an important issue to be pondered by the government in their pursuit to strengthen corporate governance practices. Although the evidence implies that Malaysian firms may generate higher efficiency through large ownership, consistent with Jensen and Meckling (1976), this result is in contrast with findings of Burkart, Gromb, Panunzi (1997) and Hill and Snell (1988) who argued that over concentration of ownership may prove to be an obstacle in exploiting growth opportunities as well as discouraging innovation and management initiative, when such situation require greater provision of capital and risk taking. Further, Shleifer and Vishny (1997) have argued that in corporate systems with a high ownership concentration, the minority shareholders may suffer risk expropriation of wealth from majority shareholders. Such expropriation merely aggravates the agency problem and reduces the firm's market value.

Through time, Malaysia's economic strength is becoming rooted in its industrial base and it is currently moving to a higher value-added service-based economy. As the Malaysian economy moves towards becoming a high-income economy, Malaysian firms face many key challenges in their quest to enhance business activities and profitability. Consequently, Malaysian firms need to be strengthened in terms of their corporate governance practices so as to ensure efficiency improvements. Their improved performance will subsequently help the country achieve high economic performance.

## 6. Conclusion

Ownership structure and firm efficiency are regarded as important fundamental issues in corporate governance. The literature on the effect of ownership structure has devoted much attention on firm performance but paid scant attention to firm efficiency. Therefore, our study of firm efficiency across different types of ownership structure attempts to fill the gap in the literature by providing new evidence on firm efficiency in Malaysia. As estimated using DEA model, this study has shown that firms with high ownership concentration appear to experience a higher efficiency compared to firms with low ownership concentration. Therefore, the government should ensure that Malaysian firms enhance their corporate governance practices in order to improve efficiency.

**Acknowledgments**

We are grateful for the financial support from Universiti Utara Malaysia under LEADS research grant 12005 for this research undertaking. We also thank all the participants in the LEADS Seminar (10 – 11th June, 2013) conducted by the Research and Innovation Management Centre (RIMC) of Universiti Utara Malaysia for their invaluable comments and support.